# CT-LungNet: A Deep Learning Framework for Precise Lung Tissue Segmentation in 3D Thoracic CT Scans

Niloufar Delfan, Hamid Abrishami Moghaddam, Mohammadreza Modaresi, Kimia Afshari, Kasra Nezamabadi, Neda Pak, Omid Ghaemi, Mohamad Forouzanfar, *Senior Member, IEEE*

*Abstract*— Segmentation of lung tissue in computed tomography (CT) images is a precursor to most pulmonary image analysis applications. Semantic segmentation methods using deep learning have exhibited top-tier performance in recent years, however designing accurate and robust segmentation models for lung tissue is challenging due to the variations in shape, size, and orientation. Additionally, medical image artifacts and noise can affect lung tissue segmentation and degrade the accuracy of downstream analysis. The practicality of current deep learning methods for lung tissue segmentation is limited as they require significant computational resources and may not be easily deployable in clinical settings. This paper presents a fully automatic method that identifies the lungs in three-dimensional (3D) pulmonary CT images using deep networks and transfer learning. We introduce (1) a novel 2.5-dimensional image representation from consecutive CT slices that succinctly represents volumetric information and (2) a U-Net architecture equipped with pre-trained InceptionV3 blocks to segment 3D CT scans while maintaining the number of learnable parameters as low as possible. Our method was quantitatively assessed using one public dataset, LUNA16, for training and testing and two public datasets, namely, VESSEL12 and CRPF, only for testing. Due to the low number of learnable parameters, our method achieved high generalizability to the unseen VESSEL12 and CRPF datasets while obtaining superior performance over Luna16 compared to existing methods (Dice coefficients of 99.7, 99.1, and 98.8 over LUNA16, VESSEL12, and CRPF datasets, respectively). We made our method publicly accessible via a graphical user interface at medvispy.ee.kntu.ac.ir.

## I. INTRODUCTION

Lung tissue segmentation is a crucial step in quantitatively analyzing thoracic computed tomography (CT) images in computer-aided medical diagnostic systems. This initial step significantly impacts the performance of downstream analyses, such as the detection of abnormalities or classification of lung diseases like cancer [1]. Automatic and accurate lung tissue segmentation is challenging due to several image characteristics, such as low contrast, dynamic range, spatial resolution, noise, and artifacts.

The lung structure is primarily made up of air, which looks dark on a CT scan. Most segmentation algorithms rely on this contrast between the lung tissue and surrounding tissues. The existing lung segmentation techniques can be categorized into three groups: basic methods, model-based methods, and deep learning methods. The first category comprises thresholding [2-4], edge detection [5], region growing [6], connected components and graph search [7], watershed [8] and their combination [9]. These algorithms separate the lungs from airway by employing a volumetric model and removing the trachea and bronchus. To fill holes and trim edges, morphological operations or other post-processing steps are applied [10].

The second category of techniques, model-based methods, have been recognized as some of the most successful lung segmentation methods [10]. They rely on a geometric or statistical atlas of the lung, which is constructed using a large number of thoracic scans. Active shape models [11-14] and atlas-based deformable models [15, 16] are some examples in which the model parameters are updated based on training images to maintain maximum compatibility with new input images. Therefore, local distortions in segmented images are less likely to occur due to the inherent knowledge provided to the model.

Deep learning architectures have emerged as a promising third category of lung segmentation methods, offering superior performance compared to model-based approaches. For instance, Alves et al. [17] proposed a fully convolutional

Niloufar Delfan, Hamid Abrishami Moghaddam, and Kimia Afshari are with Machine Vision and Medical Image Processing (MVMIP) Laboratory, Faculty of Electrical and Computer Engineering, K.N. Toosi University of Technology, Tehran, Iran (e-mail: niloufardelfan@email.kntu.ac.ir, moghaddam@kntu.ac.ir, k.afshari95@email.kntu.ac.ir).

Mohammadreza Modaresi, Neda Pak, and Omid Ghaemi are with Tehran University of Medical Sciences, Tehran, Iran (e-mail: mr-modaresi@sina.tums.ac.ir, pakneda@yahoo.com, omidghaemi@sina.tums.ac.ir)

Kasra Nezamabadi is with the Department of Computer and Information Sciences, University of Delaware, Newark, DE 19716, USA (email: kasra@udel.edu).

Mohamad Forouzanfar is with the Department of Systems Engineering, École de technologie supérieure (ÉTS), Université du Québec, Montréal, QC H3W 1W4, Canada, the Centre de recherche de l'Institut universitaire de gériatrie de Montréal (CRIUGM), Montréal, QC H3W 1W5, Canada, and the Department of Biomedical Engineering, K.N. Toosi University of Technology, Tehran, Iran (email: mohamad.forouzanfar@etsmtl.ca).

(Corresponding author: Hamid Abrishami Moghaddam.)

network (FCN) for extracting lungs from CT images. The proposed model consisted of an encoder-decoder architecture, where the encoder network used a series of convolutional and pooling layers to extract high-level features from the input image, and the decoder network used a series of transposed convolutional layers to generate the lung segmentation mask. Moreover, the authors used a conditional random fields (CRF) post-processing step to improve the spatial coherence of the lung segmentation mask generated by the FCN model and achieve more accurate segmentation results. Other recent studies have proposed innovative deep learning models that augment the popular U-Net architecture [18]. For example, Alom et al. [19] proposed a model named R2U-Net for medical image segmentation. The R2U-Net added recurrent connections and residual connections to the U-Net architecture, which allowed the model to better capture long-range dependencies in the input images. Similarly, Kadia et al. [20] proposed a 3D convolutional neural network architecture called $R^2U3D$. It incorporated recurrent connections and residual connections, which allowed the model to better capture long-range dependencies in the input images, and improve the accuracy of lung segmentation in medical images. Khanna et al. [21] modified the U-Net architecture by using residual connections in the encoder network to extract more discriminative feature representations and achieved improved performance. Gu et al. [22] presented a modified U-Net architecture called CE-Net, which utilized a context module to capture contextual information of input images and multiple branches in the decoder network to reconstruct segmentation masks at different scales. The proposed model achieved high segmentation accuracy in 2D medical images and demonstrated the effectiveness of using context information in medical image analysis.

While several studies on large datasets suggest that a deeper network can achieve a better performance in image classification and segmentation [23], computational efficiency and low parameter count are still important factors for real-time image segmentation. These considerations motivated us to explore network architectures that efficiently extract robust discriminative features. We focus on factorized convolutions and aggressive regularization.

In this study, to extract robust discriminative features from input CT images, we employed the U-Net architecture and integrated it with GoogleNet modules [24], namely InceptionV3 [25] blocks in the encoder path. For an efficient information representation from CT images, a 2.5-dimensional processing protocol was adopted [26]. CT images display image information in three dimensions. As such, considering a few consecutive slices (i.e., 2.5-dimesnion) can be more informative than individual 2-d slices for a better tissue separation and damage diagnosis. We extensively evaluated our method using three public lung tissue segmentation datasets and showed its superiority over state-of-the-art methods.

Our proposed method has been integrated in MedVisPy, a medical desktop software application written in Python, to support clinicians in processing, analyzing, and visualizing thoracic CT scans. It can be downloaded from medvispy.ee.kntu.ac.ir

## II. METHOD

### A. Datasets

Three publicly available datasets were used in this study: LUNA16, CRPF and VESSEL12. Lung Nodule Analysis 2016 (LUNA16) dataset [27] is a subset of the LIDC dataset [28] which includes 878 subjects. The images in LUNA16 represent a set of diagnostic and cancer screening lung CT scans in which the suspected lesions are annotated. The image sizes range from 512×512×95 to 512×512×733, with the voxels size of 0.78×0.78×1.25 $mm^3$. The ground truth (GT) was generated using an automatic segmentation algorithm [29]. It encompasses the right and left lungs and trachea as distinct labels.

The CRPF dataset [30] contains 50 thoracic computed tomography scans, consisting of diagnostic and screening thoracic CTs. The voxel size is 0.71×0.71×1.25 $mm^3$ and the image sizes range from 512×512×212 to 512×512×288 voxels. Two radiologists generated GTs for this study. The GT images only include the left and right lungs, not the trachea.

VESSEL12 [31], was provided for the segmentation challenge that took place in 2012 to compare methods for (semi-)automatic segmentation of the vessels in the lungs from chest computed tomography scans taken from both healthy and diseased populations with a variety of respiratory conditions that affect the lungs in a way that makes segmentation difficult. This dataset comprises CT images of 23 subjects with their corresponding lung masks, ranging in size from 512×512×355 to 512×512×543 voxels. This dataset includes diverse chest CT images, such as high resolution, low resolution, standard dose, and angio-CT.

In this study, the LUNA16 dataset was utilized for both training and testing of the proposed approach, while the VESSEL12 and CRPF datasets were solely utilized for testing purposes.

### B. Preprocessing

A two-stage preprocessing was performed before feeding CT scans to our 2.5D data generator. First, we assigned a -1000 Hounsfield units (HU) value to all boundary pixels where pixels initially have HU values of -3000. This is a standard calibration as chest CT scans typically have pixel intensities of +1000 for bone, zero for water, and -1000 for air.

Second, to remove irrelevant structures from the image and enhance the contrast, all pixel values greater than 400 were set to 400, and those less than -1000 were set to -1000. This enhanced the distinguishability of the lung tissue from its adjacent tissues. Fig. 1(a) and (b) illustrate the histogram of an image's slice from LUNA16 before and after preprocessing, respectively. As shown in Fig. 1(b), the lung tissue is.

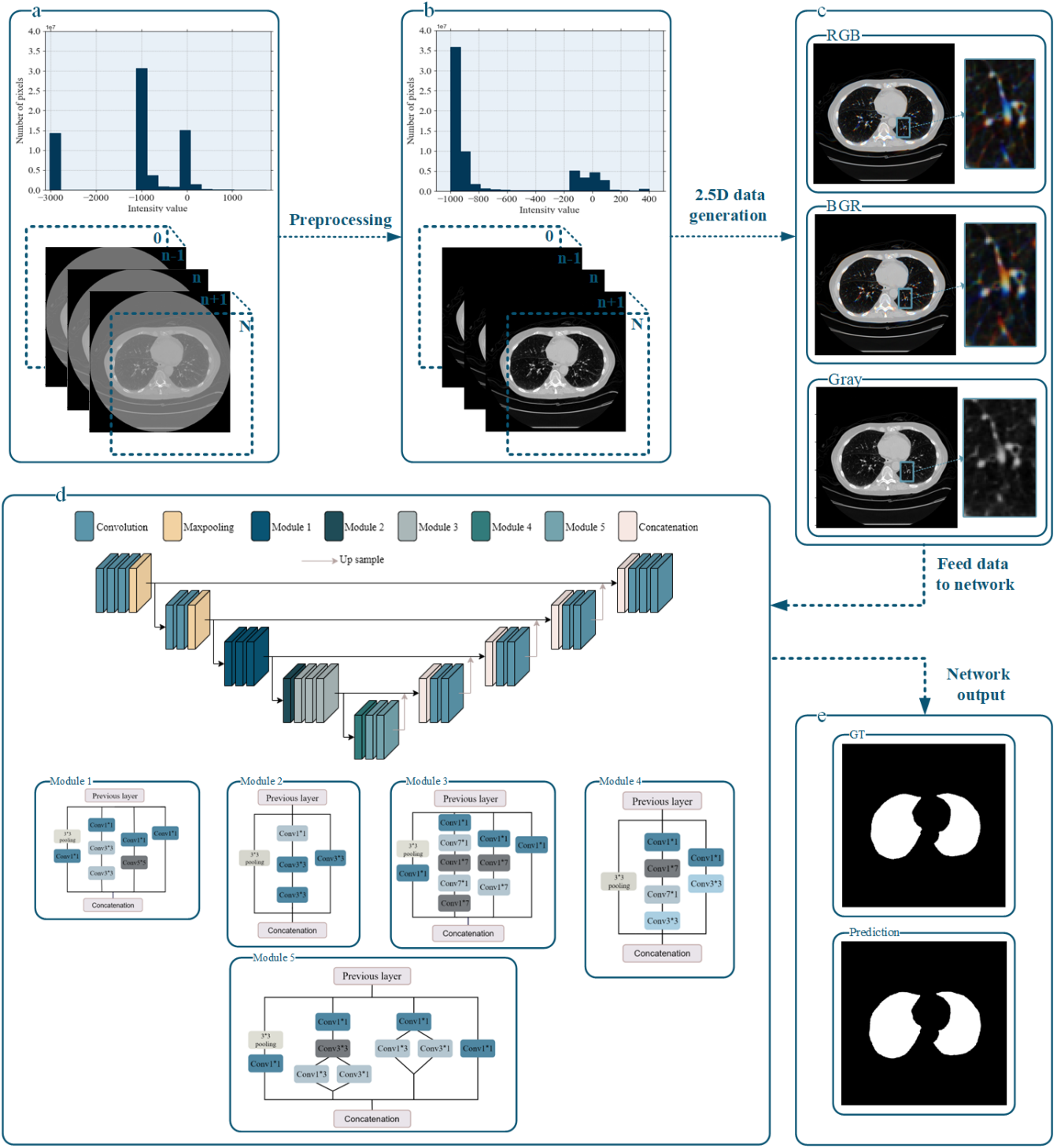

Fig. 1. An overview of the proposed method. An example of a 3D volume and its histogram (a) before (b) after HU scaling and thresholding. (c) RGB, BGR and Gray color representation of three consecutive preprocessed slices. (d) A diagram of the InceptionV3 U-Net based CNN architecture for lung segmentation from thoracic CTs. (e) An example of the network's output and its corresponding ground truth.

discernible from other tissues, and the structure of tissues remains intact

### C. 2.5-Dimensional Data Representation

After preprocessing, 2.5-dimensional images comprising n-1th, nth, and n+1th slices were generated and used to train the proposed network for segmenting the nth image slice. Inspired by [26], we generate three different data configurations by assigning $n$-1, $n$, and $n$+1 slices in different orders to the red, green, and blue network input channels, as follows (see Fig. 1c):

1. RGB configuration: slices $n − 1$, $n$, and $n + 1$ are assigned to the red, green, and blue input channels, respectively.

## TABLE I
SEGMENTATION RESULTS OF OUR METHOD ON THE LUNA16, VESSEL12, AND CRPF DATASETS

|  | Total (RGB) | Total (Gray) |
|---|---|---|
| **Luna 16** | | |
| Dice score (2D) | 0.960 | 0.953 |
| STD (2D) | 0.010 | 0.013 |
| Dice score (3D) | 0.997 | 0.992 |
| STD (3D) | 0.002 | 0.003 |
| **Vessel 12** | | |
| Dice score (2D) | 0.951 | 0.943 |
| STD (2D) | 0.005 | 0.006 |
| Dice score (3D) | 0.991 | 0.987 |
| STD (3D) | 0.002 | 0.003 |
| **CRPF** | | |
| **Radiologist 1** | | |
| Dice score (2D) | 0.944 | 0.939 |
| STD (2D) | 0.006 | 0.007 |
| Dice score (3D) | 0.989 | 0.987 |
| STD (3D) | 0.002 | 0.002 |
| **Radiologist 2** | | |
| Dice score (2D) | 0.943 | 0.938 |
| STD (2D) | 0.006 | 0.007 |
| Dice score (3D) | 0.988 | 0.987 |
| STD (3D) | 0.003 | 0.002 |

2. BGR configuration: slices n – 1, n, and n + 1 are assigned to blue, green, and red channels, respectively.

3. Grey scale configuration: slice n is assigned to all three channels.

Notably, slice n is discarded if slice n – 1 or n + 1 is outside the CT volume. All these three data structures have the same GT as the nth slice.

### D. Network Architecture

Our proposed architecture for lung tissue segmentation from 3D CT images is shown in Fig. 1(d). It is inspired by a U-Net convolutional neural network (CNN) architecture [18]. U-Net benefits from skip connections between its encoder and decoder parts, leading to better segmentation accuracy than previous networks [32].

We modified the encoder part by transfer learning. In particular, we used a pre-trained GoogleNet network with InceptionV3 [25] modules as the transferred model. We removed the fully connected (FC) layers from the InceptionV3 architecture to adopt it in the encoder part, and the decoder layers was added afterward. Our modified InceptionV3 modules receive the proposed 2.5-dimensional CT images as its three-color-channel input.

The decoder architecture is the same as that of the U-Net model. Segmentation was obtained by feeding the image to an encoder first. Each step was followed by the transfer of the resulting feature map to the corresponding layers in the decoder, followed by a concatenation of this new map with the previous decoding steps. The backpropagation of error was used to update only the decoder's weights. For each pixel, the network output is between 0 and 1. As such, a threshold value of 0.5 was used to create a binary mask. Due to the binary nature of the GT images (including lung tissue and background), the logarithmic loss function was used:

$$\text{logloss} = \frac{1}{N} \sum_{i=1}^{N} -(y_i \times \log(p_i) + (1 - y_i) \times \log(1 - p_i)) \quad (1)$$

where N is the number of pixels in the input image slice. This function compares the predicted value of all pixels ($p_i$) in the network output to their desired binary values ($y_i$) in the labelled GT image. It indicates how far the prediction is from the desired value. We used the Adam optimizer [33] via a dynamic learning rate approach with an initial rate of 0.001. The learning rate was reduced using plateau callback during the training. Once the model performance stagnated for some epochs, the callback automatically reduced the learning rate. The batch size was set to 2 and the number of training epochs was set to 50.

### III. EXPERIMENTAL RESULTS

We used the Luna16 dataset for training and testing, while VESSEL12 and CRPF datasets were used only for testing and evaluating the generalizability of our method. Over the Luna16 dataset, our method was tested using a 10-fold cross-validation, where in each fold, 10% of the training data were randomly selected as a validation set to fine-tune the hyperparameters and early stop the training to prevent overfitting.

Since the CRPF dataset does not provide ground-truth masks publicly, two radiologists, blindly to each other, annotated images in this set, and provided two sets of masks as ground-truth.

To compare our method against state-of-the-art methods, we employed the Dice coefficient, a standard assessment measure widely used in image segmentation, defined as:

$$\text{Dice} = \frac{2|S \cap G|}{|S| + |G|} \quad (2)$$

where S and G represent the predicted mask and GT, respectively, and operator ∩ indicates intersection which calculates to elements that are common to both sets [34].

The Dice coefficient was computed for every slice, and its average was calculated for all slices in a volumetric CT image. This measure was calculated for slices where at least one of the expert's annotation or automated segmentation indicates the presence of the lung; otherwise, the slice was ignored. We refer to this metric as 2D Dice score. Additionally, the 3D Dice coefficient was computed using the 3D ground-truth and 3D segmentation result obtained by concatenating segmented 2D slices.

Table I shows the performance of our method over Luna16, VESSEL12 and CRPF datasets Our method achieved an average 2D and 3D Dice coefficient of 0.96 and 0.99, respectively, on the LUNA16 dataset and 0.951 and 0.991, respectively, on the VESSEL12 dataset. Compared to the ground-truth masks annotated by the first radiologist (N. Pak) for the CRPF dataset, our method obtained an average 2D and 3D Dice score of 0.944 and 0.989, respectively. When comparing to the masks annotated by the second radiologist (O. Ghaemi), our method achieved a 2D and 3D Dice coefficient of 0.93 and 0.99, respectively. Table I shows that 2D Dice scores are generally inferior to 3D Dice scores. As the 2D Dice score is the average of the ratios between all 2D slices of a CT image, a slight difference between the ground truth and automated segmentation is more pronounced than that reflected by the 3D Dice score, which is the ratio between two volumes.

TABLE II
COMPARATIVE ANALYSIS OF THE PROPOSED METHOD VERSUS EXISTING APPROACHES.

| Author | Data size | Validation method | Features | Segmentation technique | Performance evaluation |
|---|---|---|---|---|---|
| **Private dataset** | | | | | |
| Zhou et al.[35] 2016 | 240 scans | Custom split | Automated | FCN + majority voting | 0.88 |
| Zhou et al.[36] 2021 | - | Custom split | Automated | Recurrent based on U-Net | 0.992 |
| Birbeck et al. [37] 2014 | 185 scans | Custom split | - | Statistical models +level set refinement | 0.94 |
| Sun et al. [12] 2011 | 30 scans | Custom split | - | Model based + Optimal surface finding | 0.975 |
| Gill et al. [13] 2014 | 190 scans | Custom split | 3D scale-invariant | Model based | 0.974 |
| Xu et al.[38] 2019 | 201 scans | Custom split | K-means and CNN based | k-means & CNN | 0.961 |
| **CRPF dataset** | | | | | |
| Naseri et al. [16] 2018 | 40 scans | Leave-one-out | - | Model based | 0.976 |
| Proposed method | 40 scans | Use only as test set | Automated | InceptionV3 -based U-net | **0.988** |
| **Luna16 dataset** | | | | | |
| Skourt et al. [39] 2018 | All scans | Custom split | Automated | U-Net | 0.950 |
| Alom et al. [19] 2018 | 534 scans | Custom split | Automated | Recurrent network | 0.98 |
| Gu et al. [22] 2019 | 534 scans | Custom split | Automated | Residual U-Net + RMP + DAC | 0.99 |
| Khanna et al. [21] 2020 | 50 scans | 5-fold cross-validation | Automated | Very deep Residual U-Net | 0.986 |
| Chen et al. [40] 2021 | 74 scans | Custom split | Automated | U-Net | 0.974 |
| Ma et al. [41] 2022 | 534 scans | Custom split | Automated | Context-attention network | 0.989 |
| Proposed method | 879 scans | 10-fold cross-validation | Automated | InceptionV3 -based U-net | **0.997** |
| **Vessel12 dataset** | | | | | |
| Alves et al.[17] 2018 | 20 scans | 5-fold cross validation | Automated | FCN | 0.991 |
| Khanna et al. [21] 2020 | 20 scans | 5-fold cross-validation | Automated | Very deep Residual U-Net | 0.996 |
| Soliman et al.[42] 2016 | 20 scans | Custom split | - | Markov–Gibbs random field | 0.990 |
| Chae et al. [43] 2016 | - | Custom split | Geometric | Level set method | 0.98 |
| Proposed method | 23 scans | Use only as test set | Automated | InceptionV3 -based U-net | **0.991** |

Table II compares our method against state-of-the-art methods over Luna16, VESSEL12, CRPF, and some private datasets in terms of performance, train/test methodology, feature extraction, and segmentation techniques. Over the public datasets, our method outperformed all the other methods including deep neural networks. It is worth noting that many of the methods listed in Table II were trained and evaluated using a single dataset, making it unclear how well they will perform on other datasets.

## IV. DISCUSSION

Accurate lung CT image segmentation is important for precise disease diagnosis and treatment planning, assessment of lung function, and for research purposes. It helps doctors determine the extent of disease, plan interventions such as radiation therapy and surgery, and provides valuable information for disease management. In this work, we presented a novel approach for the segmentation of lung CT images that achieved superior performance compared to state-of-the-art techniques, while maintaining a low number of learnable parameters and exhibiting strong generalizability to new data. The segmentation results reported in Table I demonstrate that our method robustly segmented CT images in the two unseen datasets, namely, VESSEL12 and CRPF. Fig. 2 illustrates the lung segmentation result obtained by applying our method to a CT image from Luna16, VESSEL2, and CRPF datasets. From left to right, three slices from the base, middle and apex sections of the CTs are shown. As illustrated, the lung regions were successfully segmented (the blue color shows True Positive segmentation) from the input CT images despite their fuzzy boundaries and irregular shapes. Moreover, Fig. 2 shows that our method remains robust when tested on unseen CT scans (from VESSEL2 and CRPF datasets). The high degree of our method's generalizability is due to our novel utilization of InceptionV3 and U-Net architectures, which provides additional feature extraction power while keeping the number of parameters as low as possible.

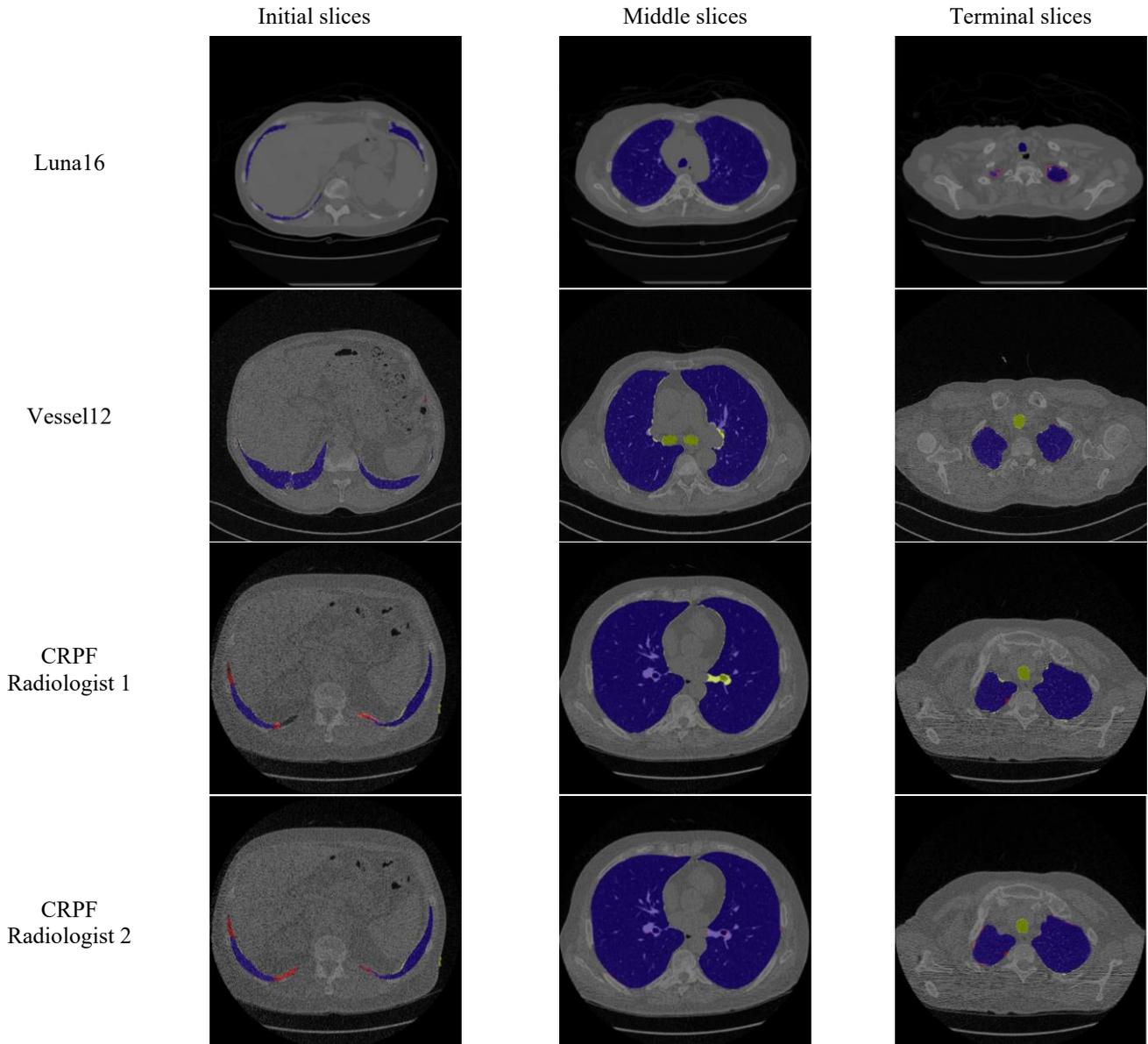

Fig. 2. Visual comparison of lung segmentation results on LUNA16, Vessel12, and CRPF Datasets. False positive (FP), false negative (FN), and true positive (TP) are indicated by red, green, and blue colors, respectively.

As shown in Table II, previous approaches often used handcrafted features [2, 42, 43], model-based methods [1, 9, 13, 16, 42], shallow networks [17, 31, 35, 38, 39] or even more complicated end-to-end CNN models [21] than ours. Model-based methods [12, 13, 16] require a significant amount of training samples to cover the diversity of the target population, thus, any uncommon lung shape needs to be added to the training set for the model to learn. In some other methods [12, 37], left and right lungs are segmented separately that can result in inconsistencies (e.g., overlap). Alves et al. [17] reported a Dice score of 99.0 for the Vessel12 dataset, but the shallow network in their model prevented the extraction of complex features for pixel-wise classification.

In some studies [19, 22, 35, 36, 38, 41], a custom train-test split method has been used for cross-validation, which could result in biased results due to differing data distributions. For example, in [21], only 50 subjects from the entire dataset were used to train, validate and test the model without specifying the selection criterion.

Moreover, most of the existing methods [16, 21, 35, 36] have reduced the resolution of lung images to 256×256 or 128×128, which can result in the generation of artifacts or even loss of important information in the images. In addition, many current methods lack full reproducibility. For instance, the method proposed by Soliman et al. [42] achieved high accuracy, but its intricate processing pipeline cannot be easily replicated without access to the source code and relevant atlases.

TABLE III
PERFORMANCE COMPARISON OF THE PROPOSED CNN MODULE VERSUS ALTERNATIVE CNNS

| Model | 3D Dice score | STD |
|---|---|---|
| FCN [44] | 0.947 | 0.004 |
| Basic U-net [18] | 0.957 | 0.004 |
| U-net+VGG16 [45] | 0.968 | 0.004 |
| U-net+VGG19 [45] | 0.970 | 0.004 |
| U-net+Resnet101 [23] | 0.973 | 0.004 |
| U-net+Resnet34 [23] | 0.978 | 0.003 |
| U-net+DenseNet [46] | 0.983 | 0.003 |
| U-net+InceptionResnet V2 [47] | 0.991 | 0.003 |
| U-net+InceptionV3 [25] | **0.997** | **0.002** |

Unlike previous approaches, our method was extensively validated using a 10-fold cross-validation procedure on Luna16 datasets, as well as on two additional datasets to evaluate its generalizability over new unseen data. In addition, it processes the high resolution 512×512 lung images without the need to reduce their resolution. Our proposed method achieved enhanced performance compared to previous approaches. Furthermore, our method has 21,787,552 learnable parameters, significantly fewer than many other deep learning architectures (e.g., number of parameters used in [21] is 25,583,592).

A comparative analysis was also carried out between our implemented model and various other CNN modules instead of InceptionV3 for semantic segmentation. The results are summarized in Table IV. Our proposed model was also compared with state-of-the-art segmentation methods, including FCN [44], U-Net [18], and a modified version of U-Net. The Luna16 dataset with similar training and testing strategies was used for all these networks. As reported in Table IV, the FCN model achieved the lowest performance in terms of Dice score. Compared to other models, FCN has a very shallow architecture, which justifies its low performance. The U-Net architecture has performed better than the FCN model with a Dice score of 95.7. Accordingly, we conclude that the U-Net model is a better candidate for lung segmentation. Nevertheless, the performance degradation problem becomes apparent as the network becomes deeper.

## V. CONCLUSION

In this study, we introduced a fully automated framework for lung CT image segmentation using a deep convolutional neural network. Our framework combined the advantages of both InceptionV3 and U-Net architectures to produce a high-performing network. To further improve the generalization and feature extraction capabilities, a 2.5D CT representation was employed. The results showed that our method outperformed other techniques in lung segmentation, even in complex lung regions, when evaluated on three benchmark datasets, LUNA16, VESSEL12, and CRPF. Our approach provides a promising solution for precise and efficient lung CT segmentation.

Our proposed approach has the potential to contribute to the accurate segmentation of small nodules in lung CT images, which is crucial for enhancing lung cancer diagnosis and treatment. Future research can focus on the application of our method to detect and segment small nodules in lung CT images, facilitating early detection and treatment of lung cancer. This can potentially lead to improved treatment outcomes and increased chances of survival for patients with lung cancer.

Although our study focused primarily on lung segmentation, our proposed method has the potential to be applied to a wide range of medical imaging tasks beyond lung segmentation. The combination of InceptionV3 and U-Net architectures, along with the 2.5D CT representation, can be utilized for the segmentation of other organs such as the liver, heart, or kidneys. The successful application of our method to these organs could lead to improved accuracy and efficiency in medical imaging diagnosis and treatment planning. Additionally, the proposed method could be adapted to segment other types of abnormalities or structures in medical images, such as tumors, blood vessels, or bones, for various medical applications. Therefore, our proposed method has great potential for advancing medical imaging analysis and diagnosis beyond lung CT segmentation.